\documentclass[12pt]{article}  
\usepackage{a41}
\usepackage{epsfig}
\newcommand{\gap}{\stackrel{>}{\sim}}
\newcommand{\lap}{\stackrel{<}{\sim}}

\newcommand{\lsim}{\stackrel{<}{\sim}}
\def \bfk{\mbox{\boldmath $k_\perp$}}

\begin{document}

\vspace*{1.0cm}

August 1996  \hfill  DESY 96--128  \\
\vspace{-0.2ex}
             \hfill  INFNCA--TH~9616  \\

 \hfill  hep--ph 9608xxx \\

\vspace*{1.5cm}

\begin{center}  
   \begin{Large} \begin{bf}
{On the Physics Potential of      \\
Polarized Nucleon--Nucleon Collisions at HERA
\footnote{Contribution to the Proceedings of the Workshop
   on `Future Physics at HERA', ~~DESY--Hamburg, \\
 \hspace*{3ex} Oct. 1995 - May 1996}
}  \\
   \end{bf} \end{Large} 

   \vspace*{20mm}
   \begin{large}
M. Anselmino$^a$, E. Andreeva$^b$, V. Korotkov$^c$, F. Murgia$^d$, \\
W.--D. Nowak$^{e,}\footnote{e-mail: nowakw@ifh.de
}$, S. Nurushev$^c$, O. Teryaev$^f$, A. Tkabladze$^f$ \\
\end{large}

   \vspace*{15mm}
$^a$ University of Torino, Italy; 
$^b$ MEPHI Moscow, Russia; 
$^c$ IHEP Protvino, Russia; \\
$^d$ University of Cagliari, Italy;
$^e$ DESY-IfH Zeuthen, Germany;
$^f$ JINR Dubna, Russia
\end{center}

\vspace{2.0cm}

\begin{abstract}
The physics of polarized nucleon--nucleon collisions originating 
from an internal polarized target in the HERA proton beam is 
investigated. 
Based on 240 pb$^{-1}$ integrated luminosity at 40 GeV c.m. energy, 
statistical 
sensitivities are given over a wide $(x_F, p_T)$--range for a variety 
of inclusive and exclusive final states. By measuring 
single spin asymmetries unique information can be obtained on 
higher twist contributions and their $p_T$-dependence. 
From double spin asymmetries in both photon and J/$\psi$ production
it appears possible to measure the 
polarized gluon distribution in the range 0.1~$\leq x_{gluon} 
\leq$~0.4 with a good statistical accuracy.
\end{abstract}

\newpage

%
\section{Introduction}
%

There is a widespread general consensus, much grown in the last years
and based both on surprising experimental results and intense theoretical
activity, that the spin structure and the spin dynamical properties
of nucleons and hadrons in general are far from being understood; a 
satisfactory knowledge of hadronic structure and dynamics and their correct 
description in terms of constituents cannot ignore the spin subtleties and 
more experimental information is badly required. \\
An experiment (`{\it HERA--$\vec{N}$}' \cite{desy96-095})
utilising an internal polarized nucleon target in the 820~GeV 
HERA proton beam would constitute a natural extension of the studies of the 
nucleon spin structure in progress at DESY with the {\it HERMES}
experiment \cite{her1}.
Conceivably, this would be the only place where to study high energy 
nucleon--nucleon spin physics besides the dedicated RHIC spin program at BNL 
\cite{RSC} supposed to start early in the next decade. \\
An internal polarized nucleon target offering unique features such as 
polarization above 80\% and no or small dilution, can be safely operated
in a proton ring at high densities up to $10^{14}$ atoms/cm$^2$ \cite{ste1}.
As long as the polarized target is used in conjunction with an unpolarized proton
beam, the physics scope of {\it HERA--$\vec{N}$} would be focused to 
`phase~I', i.e. measurements of single spin asymmetries.
Once later polarized protons should become available, the same set-up would 
be readily available to measure a variety of double spin asymmetries. 
These `phase~II' measurements would constitute an alternative 
--~fixed target~-- approach to similar physics which will be accessible to 
the collider experiments {\it STAR} and {\it PHENIX} at the low end of 
the RHIC energy scale ($\sqrt{s}~\simeq~50$~GeV) \cite{bun1}.

We shall briefly discuss here the physics motivations to measure
single and double spin asymmetries in several $p \vec N$ inclusive and
exclusive processes; more details and further discussions, together
with a complete list of references, which we cannot give here for lack
of space, can be found in Ref. \cite{desy96-04}. We recall that the 
integrated luminosity calculation is based upon realistic figures. 
For the average beam and target polarisation
$P_B = 0.6$ and $P_T = 0.8$ are assumed, respectively. A combined trigger  and 
reconstruction efficiency of $C \simeq 50\%$ is anticipated.
Using $\bar{I}_B = 80 \; \mbox{mA} = 0.5 \cdot 10^{18} \; s^{-1}$
for the average HERA proton beam current (50\% of the design value)
and a rather conservative
polarized target density of $n_T = 3 \cdot 10^{13}$ atoms/cm$^2$ 
the projected integrated luminosity becomes
${\cal{L}} \cdot T = 240 \; pb^{-1}$
when for the total running time $T$ an equivalent of $T = 1.6 \cdot 10^7 \;s$ at 
100~\% efficiency is assumed. This corresponds to about 3 real years under
present HERA conditions. However,
proton currents much higher than the original HERA design 
value of 160 mA are envisaged in the HERA luminosity upgrade program.
In addition, 
experience from UA6 running at CERN shows that after having gained 
some practical running experience it presumably becomes
feasible to operate the polarized gas target at about 3 times 
higher density without seriously affecting the proton beam lifetime. 
Hence in a few years 500 pb$^{-1}$ {\it per year} will presumably
become a realistic figure. \\
We note that, except in the case of single spin asymmetries, we took
into account major acceptance limitations and jet detection efficiencies.
Hence it can be anticipated that the
sensitivities shown in the rest of the paper are
realistic for an about one year's running of a
future polarized nucleon--nucleon scattering experiment at HERA.
Any better estimate requires considerably
intensified efforts to be invested along many different directions, like
machine and target limitations, detector capabilities versus rate,
acceptance, costs etc.

\newpage

\section{Single Spin Asymmetries}
%

Single spin asymmetries in large $p_T$ inclusive production, both 
in proton-nucleon and lepton-nucleon interactions, have recently 
received much attention (for references see \cite{desy96-04}).
The naive expectation that they should be zero in perturbative QCD has 
been proven to be false, both experimentally and theoretically.
It is now clear that higher twist effects are responsible for 
these asymmetries, which should be zero only in leading twist-2 
perturbative QCD.  \\
Several models and theoretical analyses suggest possible higher
twist effects: there might be twist-3 dynamical contributions, which we 
shall denote as hard scattering higher twists; there might also be 
intrinsic $k_\perp$ effects, both in the quark fragmentation process 
and in the quark distribution functions. The latter are not by 
themselves higher twist contributions - they are rather non-perturbative 
universal nucleon properties - but give rise to twist-3 contributions
when convoluted with the hard scattering cross sections.
The dynamical contributions result from a short distance part
calculable in perturbative QCD with slightly modified Feynman 
rules, combined with a long distance part related to quark-gluon 
correlations. \\

An intrinsic $k_\perp$ effect in the quark fragmentation
is known as Collins or
sheared jet effect; it simply amounts to say that the number 
of hadrons $h$ (say, pions) resulting from the fragmentation
of a transversely polarized quark, with longitudinal momentum
fraction $z$ and transverse momentum $\bfk$, depends on the 
quark spin orientation. That is, one expects the {\it quark 
fragmentation analysing power} $A_q(\bfk)$ to be different from zero:

\vspace{+1ex}
\begin{equation}
\label{qfrana}
 A_q(\bfk) \equiv 
{D_{h/q^\uparrow}(z, \bfk) - D_{h/q^\downarrow}(z, \bfk)
\over 
D_{h/q^\uparrow}(z, \bfk) + D_{h/q^\downarrow}(z, \bfk)}
\not= 0 
\end{equation}
\vspace{+1ex}

where, by parity invariance, the quark spin should be orthogonal
to the $q-h$ plane. Notice also that time reversal invariance 
does not forbid such quantity to be $\not= 0$ because of the 
(necessary) soft interactions of the fragmenting quark with
external strong fields, i.e. because of final state interactions.
This idea has been applied to the computation of the single 
spin asymmetries observed in $pp^\uparrow \to \pi X$ \cite{art}. \\

A similar idea applies to the distribution functions, provided 
soft gluon interactions between initial state partons are present and 
taken into account, which most certainly is the case for hadron-hadron 
interactions. That is, one can expect that the number of quarks with 
longitudinal momentum fraction $x$ and transverse intrinsic motion 
$\bfk$ depends on the transerve spin direction of the parent nucleon,
so that the {\it quark distribution analysing power} 
$N_q(\bfk)$ can be different from zero:

\vspace{+1ex}
\begin{equation}
\label{qdisana}
 N_q(\bfk) \equiv 
{f_{q/N^\uparrow}(x, \bfk) - f_{q/N^\downarrow}(x, \bfk)
\over 
f_{q/N^\uparrow}(x, \bfk) + f_{q/N^\downarrow}(x, \bfk)}
\not= 0 
\end{equation}
\vspace{+1ex}

This effect also has been used to explain the single 
spin asymmetries observed in $pp^\uparrow \to \pi X$ \cite{ans}.
Note that both $A_q(\bfk)$ and $N_q(\bfk)$
are leading twist quantities which, when convoluted 
with the elementary cross-sections and integrated over $\bfk$, give
twist-3 contributions to the single spin asymmetries.  \\

\newpage

Each of the above mechanisms might be present and might be important
in understanding twist-3 contributions; in particular the quark
fragmentation or distribution analysing powers look like new 
non-perturbative universal quantities, crucial in clarifying the 
spin structure of nucleons. It is then of great importance to 
study possible ways of disentangling these different contributions
in order to be able to assess the importance of each of them.
We propose here to measure the single spin asymmetry

\vspace{+1ex}
\begin{equation}
\label{ssasym}
 {d\sigma^{AB^\uparrow \to CX} - d\sigma^{AB^\downarrow \to CX}
\over d\sigma^{AB^\uparrow \to CX} + d\sigma^{AB^\downarrow \to CX}}\, 
\end{equation}
\vspace{+1ex}

in several different processes $A B^\uparrow \to C X$ which 
should allow to fulfil such a task. To obtain a complete picture we need 
to consider nucleon-nucleon interactions together with other processes, 
like lepton-nucleon scattering, which might add valuable information.
For each of them we discuss the possible sources of higher twist
contributions, distinguishing, according to the above discussion, 
between those originating from the hard scattering and those originating
either from the quark fragmentation or distribution analysing power.

\vspace{+1ex}
$\bullet \quad pN^\uparrow \to hX$ \\
In this process all kinds of higher twist contributions may be present;
this asymmetry {\it alone} could not help in evaluating the relative 
importance of the different terms.

\vspace{+1ex}
$\bullet \quad pN^\uparrow \to \gamma X, \> pN^\uparrow \to \mu^+\mu^- X,
\> pN^\uparrow \to jets + X$  \\
Here there is no fragmentation process, and we remain with possible 
sources of non-zero single spin asymmetries in the hard scattering 
or the quark distribution analysing power.

\vspace{+1ex}
$\bullet \quad lN^\uparrow \to hX$ \\
In such a process the single spin asymmetry can originate either from 
hard scattering or from $k_\perp$ effects in the fragmentation function, 
but not in the distribution functions, as soft initial state interactions
are suppressed by powers of $\alpha_{em}$. Moreover, this  process allows, 
in principle, a direct measurement of the Collins effect, i.e. of 
the quark fragmentation analysing power, via a measurement of the leading-twist
difference of cross sections for the production of two identical
particles inside the same jet, with opposite $\bfk$.

\vspace{+1ex}
$\bullet \quad
lN^\uparrow \to \gamma X, \> \gamma N^\uparrow \to \gamma X, \>
lN^\uparrow \to \mu^+\mu^- X, \> lN^\uparrow \to jets + X$   \\
Each of these processes yields a single spin asymmetry which cannot 
originate from distribution or fragmentation $k_{\perp}$ effects;
it may only be due to higher-twist hard scattering effects, which would  
thus be isolated.

\vspace{+1ex}
It is clear from the above discussion that a careful and complete
study of single spin asymmetries in several processes might be a 
unique way of understanding the origin and importance of higher twist
contributions in inclusive hadronic interactions; not only, but 
it might also allow a determination of  fundamental non-perturbative
properties of quarks inside polarized nucleons and of polarized 
quark fragmentations. Such properties should be of universal value 
and applicability and their knowledge might be as important as the 
knowledge of unpolarized distribution and fragmentation functions.    \\

In the following we discuss the capability of HERA-$\vec N$ to investigate
some of these processes. \\

\newpage

\begin{figure}[h]
\centering
\epsfig{file=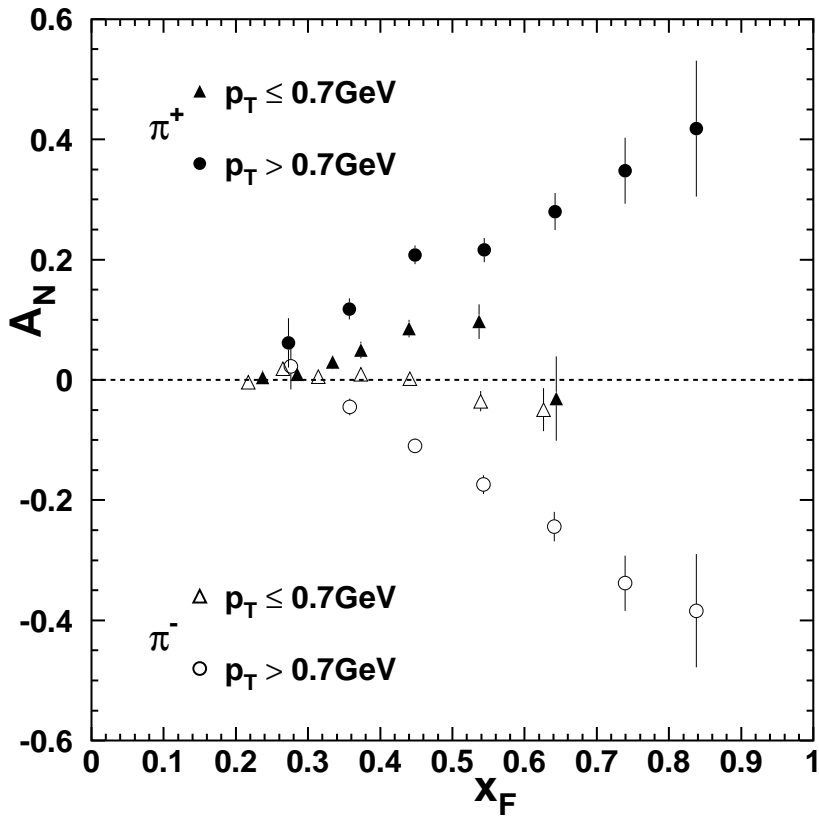, width=12cm}
\caption{\it Single spin asymmetry in inclusive pion production
         $p^{\uparrow}~+~p~\rightarrow~\pi^{0\pm}~+~X$ measured by
         the E704 Collaboration \protect \cite{704pi} 
         and shown for two subregions of $p_T$. }
         \label{e704data}
\end{figure}

{\bf Inclusive pion production} $p^{\uparrow} p \rightarrow \pi^{0\pm}X$
at 200 GeV exhibits surprisingly large single spin 
asymmetries, as it was measured a few years ago
by the E704 Collaboration using a transversely 
polarized beam \cite{704pi}. For any kind of pions the asymmetry $A_N$ shows a
considerable rise above 
$x_F > 0.3$, i.e. in the fragmentation region of
the polarized nucleon. It is positive for both
$\pi^+$ and $\pi^0$ mesons, while it has the opposite sign for $\pi^-$ mesons.
The charged pion data taken in the $0.2 < p_T < 2$~GeV range 
were split into two samples at $p_T$~=~0.7~ GeV/c; the observed rise
 is stronger for the high $p_T$ sample, as can be
seen from fig.~\ref{e704data}.

\begin{figure}[ht]
\centering
\epsfig{file=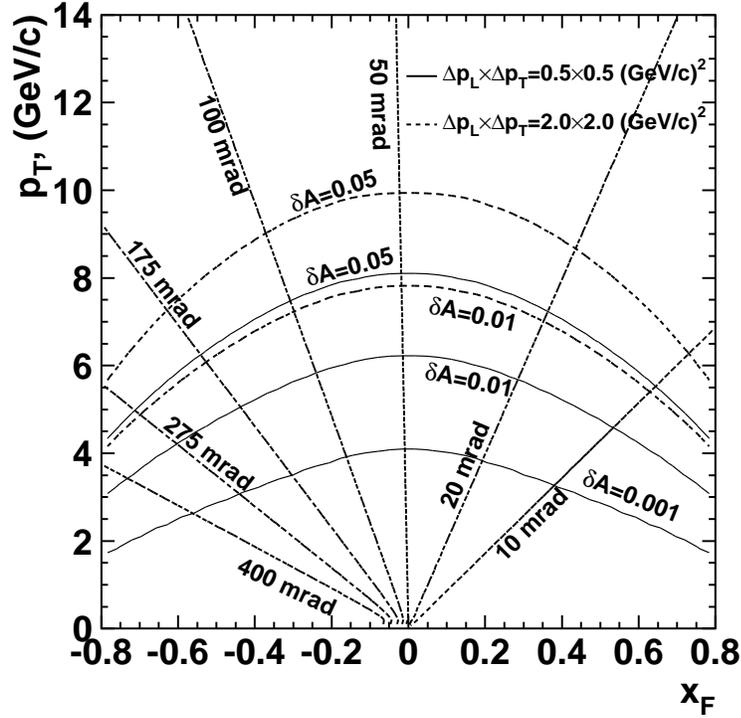,width=10.5cm}
\caption{\it Contours of the asymmetry sensitivity levels for $\pi^+$
       production in the ($p_T,~x_F$) plane. Lines of  constant 
       laboratory angles of the pion are shown.
        }     \label{fpisens}
\end{figure}

\begin{figure}[hb]
\vspace*{-10mm}
\centering
\epsfig{file=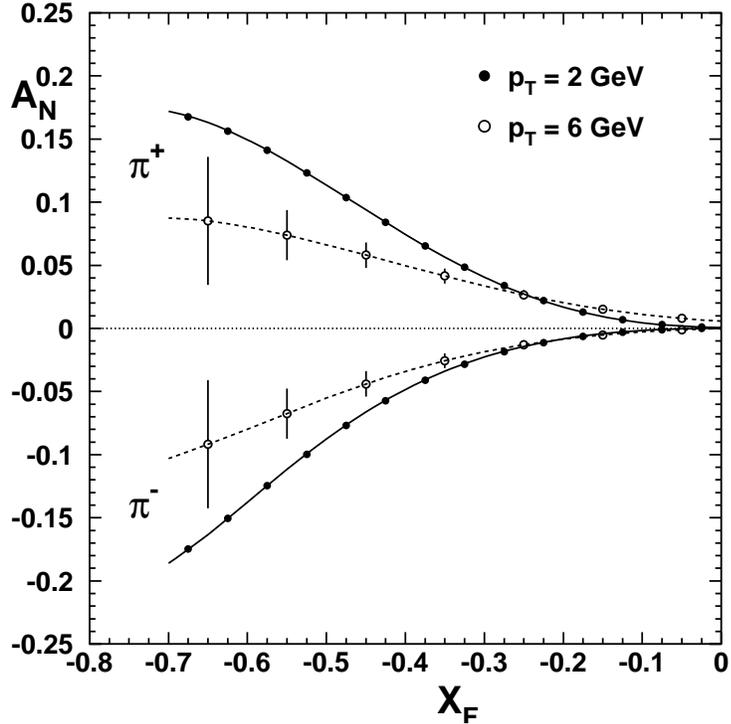,width=10.5cm}
\caption{\it Capability of HERA-$\vec{N}$ to discriminate predictions
             for different $p_T$.        } 
             \label{asymurgia}
\end{figure}

Contours characterizing different HERA-$\vec{N}$ sensitivity levels 
($\delta A_N = 0.001$, $0.01$ and $0.05$) for an asymmetry measurements 
in the reaction  $pp^{\uparrow} \rightarrow \pi^+ X$ are shown in
fig.~\ref{fpisens}.  
Note that in the large $p_T$ region the contours calculated with big
$\Delta p_L \times \Delta p_T$ bins are appropriate, since usually a larger
bin size is chosen where the statistics  starts to decrease.
We can conclude that the accessible $p_T$ values are significantly larger
than those E704 had; the combined $p_T$ dependence of all involved 
higher-twist
 effects can be measured with good accuracy ($\delta A_N \leq 0.05$)
up to transverse momenta of about 10~GeV/c in the central region $|x_F| < 0.2$ 
and up to 6~GeV/c in the target fragmentation region.
This corresponds to
an almost one order of magnitude extension in the $p_T$ range in comparison
to E704. The capability of HERA-$\vec{N}$ to really prove a predicted $p_T$
dependence is shown in fig.~\ref{asymurgia}, where the curve was obtained 
assuming a non--zero quark distribution analysing power 
(cf. eq.~\ref{qdisana}) according to Ref. \cite{ans}.  \\

\newpage

{\bf Inclusive direct photon production}, $p p^{\uparrow} \to \gamma X$,
proceeds without fragmentation, i.e. the photon carries directly the 
information from the hard scattering process. Hence this process measures 
a combination of initial $\bfk$ effects and hard scattering twist--3 
processes. The first and only results up to now were obtained by E704 
Collaboration \cite{Phot704} 
showing an asymmetry compatible with zero within 
large errors for $2.5 < p_T <3.1$~GeV/c in the central region 
$ | x_F | \lsim 0.15$.   \\
The experimental sensitivity of HERA-$\vec N$ 
was determined using PYTHIA~5.7 by
simultaneous simulation of the two dominant hard subprocesses contributing 
to direct photon production, i.e. gluon--Compton scattering
($qg \rightarrow \gamma q$) and quark--antiquark annihilation
($q \bar q \rightarrow \gamma g$),
and of background photons that originate mainly from $\pi^0$ and $\eta$ decays. 
It turns out that a good sensitivity (about 0.05) 
can be maintained up to $p_T \leq$ 8 GeV/c. 
For increasing transverse momentum the annihilation subprocess and the
background photons are becoming less essential;
we expect to be able to detect a clear dependence on $p_T$,
of the direct photon single spin asymmetry. \\

{\bf Inclusive J$/\psi$ production} was 
calculated in the framework of the colour singlet model \cite{psicosi}.
Our calculations at HERA-$\vec N$ energies \cite{desy96-04}
show an asymmetry less than 0.01 in the region $|x_F|~<~0.6$,
i.e. the effect is practically unobservable. \\
%

\section{Double Spin Asymmetries}
%

Perturbative QCD allows sizeable lowest order double spin asymmetries 
for various 2$\rightarrow$2 partonic subprocesses. Relying on 
the factorization theorem a rich spectrum of 
asymmetries at the hadronic level can be predicted
which constitute the backbone of the RHIC spin physics program. 
When both incoming particles are longitudinally polarized, the
insufficient knowledge of the polarized gluon distribution makes the 
predictions for double spin asymmetries $A_{LL}$ to some extent uncertain. 
Conversely, the measurement of $A_{LL}$ in certain final states seems to be 
one of the most valuable tools to measure the polarized gluon distribution
function in the nucleon. The presently most accurate way to do so
is the study of those processes which can be calculated 
in the framework of perturbative QCD, i.e. for which the involved 
production cross sections and subprocess asymmetries can be predicted. 
Both {\it direct photon (plus jet)} and 
{\it J/$\psi$ (plus jet)} production are most 
suited because there are only small uncertainties due to fragmentation. \\
In the following we discuss the corresponding capabilities of 
HERA--$\vec{N}$, operated in doubly polarized mode (`Phase~II'), 
to perform such measurements.  \\

{\bf Inclusive photon production} with HERA--$\vec{N}$ was the subject of
a very recent study \cite{GorVog}. Basing on a NLO calculation
\begin{figure}
\vspace*{-10mm}
\centering
\epsfig{file=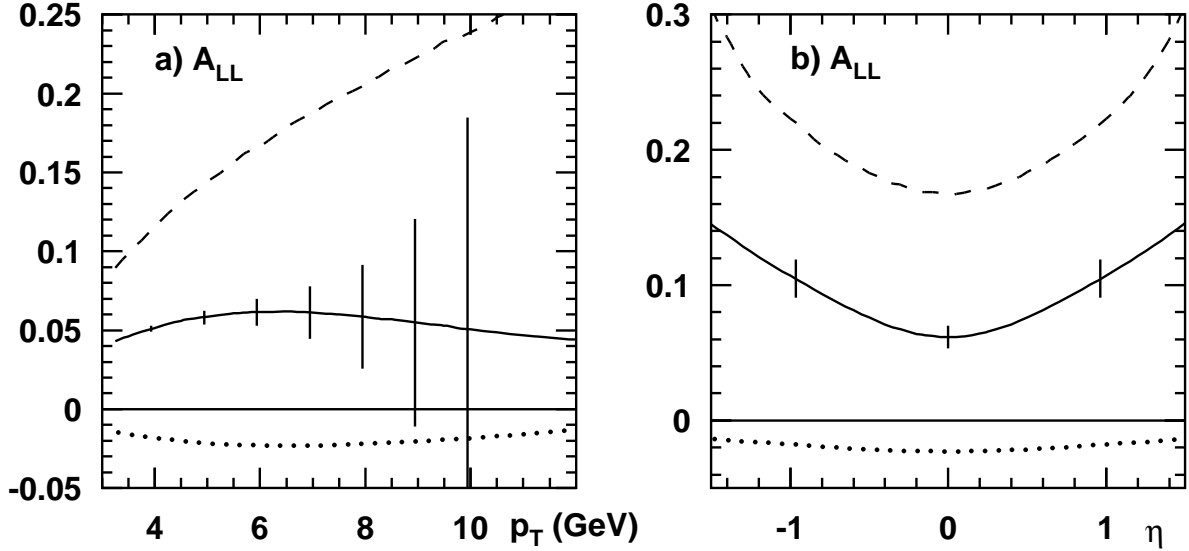,width=18cm}
\caption{\it Inclusive photon production: 
Double spin asymmetry vs. a) $p_T$ and b) $\eta$ for the
NLO 'valence set' of Ref. \protect\cite{GorVog} (full line), 
shown in conjunction with the
HERA--$\vec{N}$ statistical sensitivity. The dotted line corresponds to
set C of Ref. \protect\cite{GehrStir} and the dashed line is close to 
set A of Ref. \protect\cite{GehrStir_LO}
                                  }     \label{figvogel}
\end{figure}
rather firm predictions were obtained including an
assessment of their theoretical uncertainties; the latter turned out
to be of rather moderate size. In fig. \ref{figvogel} three different 
predictions for the asymmetry are shown  
in dependence on $p_T$ and pseudorapidity $\eta$, in conjunction with
the attainable statistical uncertainty of HERA--$\vec{N}$. We note that 
the dashed line is rather close to the prediction of Ref. 
\cite{GehrStir_LO}, set A. As can be seen, there is sufficient
statistical accuracy up to transverse momenta of about 8 GeV/c
to discriminate between different polarized gluon distribution functions
(cf. fig \ref{figvogel}a). At $p_T$~=~6~GeV/c there is sufficient accuracy
to check the asymmetry prediction for photon pseudorapidities between
-1.5 and 1.5 (cf. fig \ref{figvogel}b). \\

\newpage

In {\bf photon plus jet production} the away-side jet is measured as well
and the complete kinematics of the 
2$\rightarrow$2 subprocess can be reconstructed. In this
case the asymmetry $A_{LL}$ can be directly related to the polarized
gluon distribution if a certain subprocess can be selected. Using this
approach {\it photon plus jet} production was discussed in 
Ref. \cite{desy96-04} as a tool to directly measure $\Delta$G/G. 
The quark--antiquark annihilation subprocess is suppressed relatively to
quark-gluon Compton scattering because of the lower density of antiquarks 
(of the polarized sea) compared to gluons (polarized gluons). 
The absolute statistical error of $\Delta G(x_g) / G(x_g)$ was 
obtained as

\begin{eqnarray}
\label{err}
\delta [{{\Delta G(x_g)}\over {G(x_g)}} ]=
{{\delta A_{LL}}\over {A_{DIS} \cdot \hat a_{LL}}}.
\end{eqnarray}

\begin{figure}[ht]
\vspace*{-5mm}
\centering
\epsfig{file=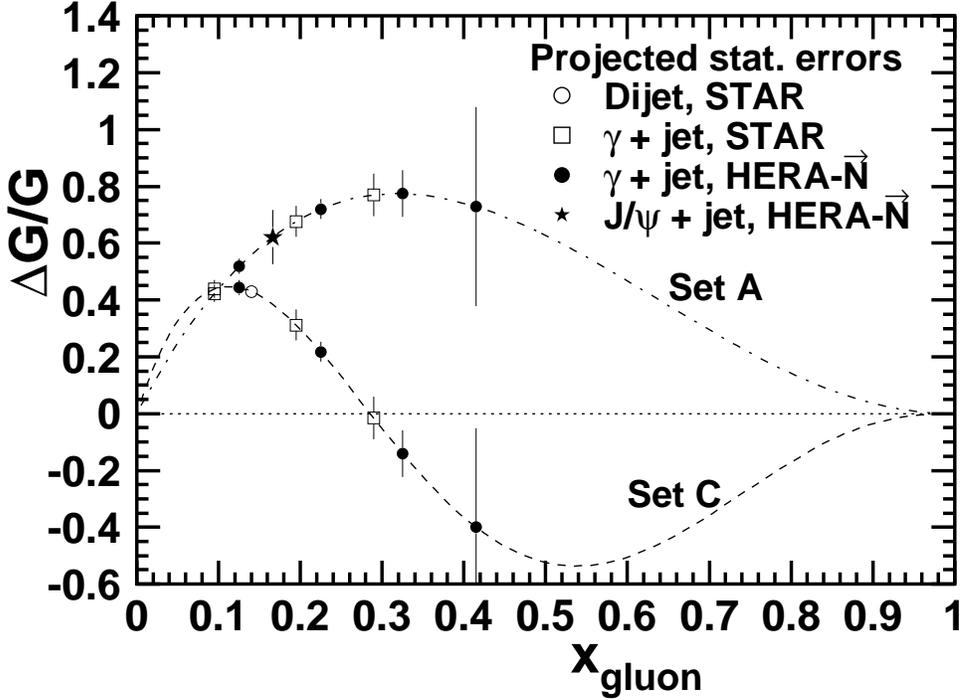,width=14.5cm}
\caption{\it Typical predictions for the polarized gluon distribution confronted
       with the projected statistical errors expected for
       HERA-$\vec{N}$ and RHIC experiments.  }    \label{photonjet}
\end{figure}

Here $A_{DIS}$ and $\hat a_{LL}$ are to be taken at the appropriate values of 
$x_F$ and $x_g$, respectively. In calculating the r.h.s. of eq.~\ref{err} we
had to take into account the
influence of the acceptance, as described in \cite{desy96-04}. \\
In fig. \ref{photonjet} the calculated HERA--$\vec{N}$ statistical sensitivity, 
on the present level of understanding, is shown vs. $x_{gluon}$ in conjunction
with predicted errors for {\it STAR} running at RHIC at 200 GeV c.m. energy
\cite{yok3}. 
The errors demonstrate clearly that in the region  
$0.1 \leq x_{g} \leq 0.4$ a significant result can be expected.
This statement will very probably remain valid if once the
systematic errors will have been estimated. As can be seen, the measurement
of $\Delta G / G $ from {\it photon plus jet} production
in doubly polarized nucleon-nucleon collisions
at HERA can be presumably performed with an accuracy being about competitive 
to that predicted for RHIC.  \\
At this point we note that
the HERA--$\vec{N}$ fixed target kinematics causes additional problems
for the jet reconstruction when compared to a collider experiment. As
obtained from rather preliminary investigations \cite{desy96-04}, the
number of photon events accompanied by a successfully reconstructed jet
decreases considerably when approaching
lower values of $p_T$ and, correspondingly, of $x_{gluon}$. To have a more
realistic statistical significance than that shown in Ref. \cite{desy96-04}
we now include these preliminary jet reconstruction efficiencies. As a
result, the statistical error bars became somewhat larger for smaller
transverse momenta. \\

{\bf $J/\psi$ Production.} Compared to direct photon production the 
production of quarkonium
states below the open charm threshold is a similarly clean tool to
measure the polarized gluon distribution. Hence many  statements made
in the previous section apply here, as well, and the principle of analysis is
very similar. Because of the relatively large quark mass the $c\bar c$
production cross section and the expected asymmetry are supposed to be
calculable perturbatively.  \\
 Quarkonium production has traditionally been calculated in the 
color-singlet model (CSM) \cite{CSM} where the quark-antiquark
pair  is produced in a color-singlet state with the quantum numbers of
the corresponding hadron. This  heavy mass pair then creates the hadronic
state with a probability determined by the appropriate quarkonium wave
function  at the origin. It is assumed that for heavy
quarks soft gluon emission is negligible, as also
other non-perturbative effects like higher twist contributions.
 While this model gives a reasonable
description of $J/\psi$ production cross section  shapes over
 $p_T$ and $x_F$, it completely fails in the explanation of the integrated 
cross section; a K factor of $7~\div~10$
 is needed to explain the data.
The anomalously large cross section \cite{CDF} for $J/\psi$ 
production at large transverse
momenta found at the Tevatron reveals another bad feature of the CSM;
it is not able to explain the
large $\psi'$  and direct $J/\psi$ production rates at CDF.
All these observations  led to the understanding that fragmentation
and hadronization of color-octet \cite{octet} {\it $q\bar q$}
pairs are essential in the heavy quarkonium production process. \\
Within the framework of the color-octet mechanism 
the quarkonium production process can be
separated, according to the factorization hypothesis, into a
short and a long distance part.
 The former describes the quark-antiquark pair production at
 small distances and can be computed perturbatively. The latter
 is responsible for the creation of a particular hadronic state 
 from the quark-antiquark pair; its matrix elements can not be
 calculated perturbatively.   \\
 The shapes of the $p_T$ distribution of short distance 
 matrix elements calculated within the color-octet model 
indicate that the new mechanism 
seems to be able to explain the Tevatron data for
direct $J/\psi$ and $\psi'$ production at large $p_T$ \cite{CL}.
We note that
unlike color-singlet matrix elements being connected to the subsequent
hadronic non-relativistic wave functions at the origin, color-octet long
distance matrix elements are unknown and have to be extracted from the
experiment. 
As it was  shown recently \cite{hadro,MYprod} the color-octet 
mechanism gives the dominant contribution in 
$J/\psi$-production at HERA--$\vec{N}$ energies. 
This concerns not only
the total cross section \cite{hadro}, but also $J/\psi$ production
at non-zero $p_T$, i.e. at those values of transverse momenta which are
not due to internal motion of partons inside the colliding hadrons,
$p_T~\gap~1.5$ GeV/c \cite{MYprod}.  \\

{\bf For inclusive $J/\psi$ production} we calculated the double spin-asymmetry
  within the framework of the color-octet model using the value of
the color-octet matrix elements from \cite{MYprod}. It is interesting
to note that the magnitude of the asymmetry does almost not depend
on the choice
of the color-octet matrix elements, if one assumes that the transition matrix
  elements of  octet $^3P_J$-states into $J/\psi$
  are negligible  compared to those of the $^1S_0$ state.
This is the most likely scenario since only
in this case it is possible to establish a consistence between different
combinations of the above two
matrix elements, as extracted from CDF data  \cite{CL}  and
photo- \cite{photo} or hadroproduction \cite{hadro}
data. We underline that
the measurement of the double spin asymmetry in $J/\psi$
 production would allow to 
 extract information about the color-octet matrix elements {\it separately},
whereas  from unpolarized experiments it is only possible to extract 
combinations of them.
\begin{figure}[h]
\centering
\begin{minipage}[c]{8.2cm}
\centering
\epsfig{file=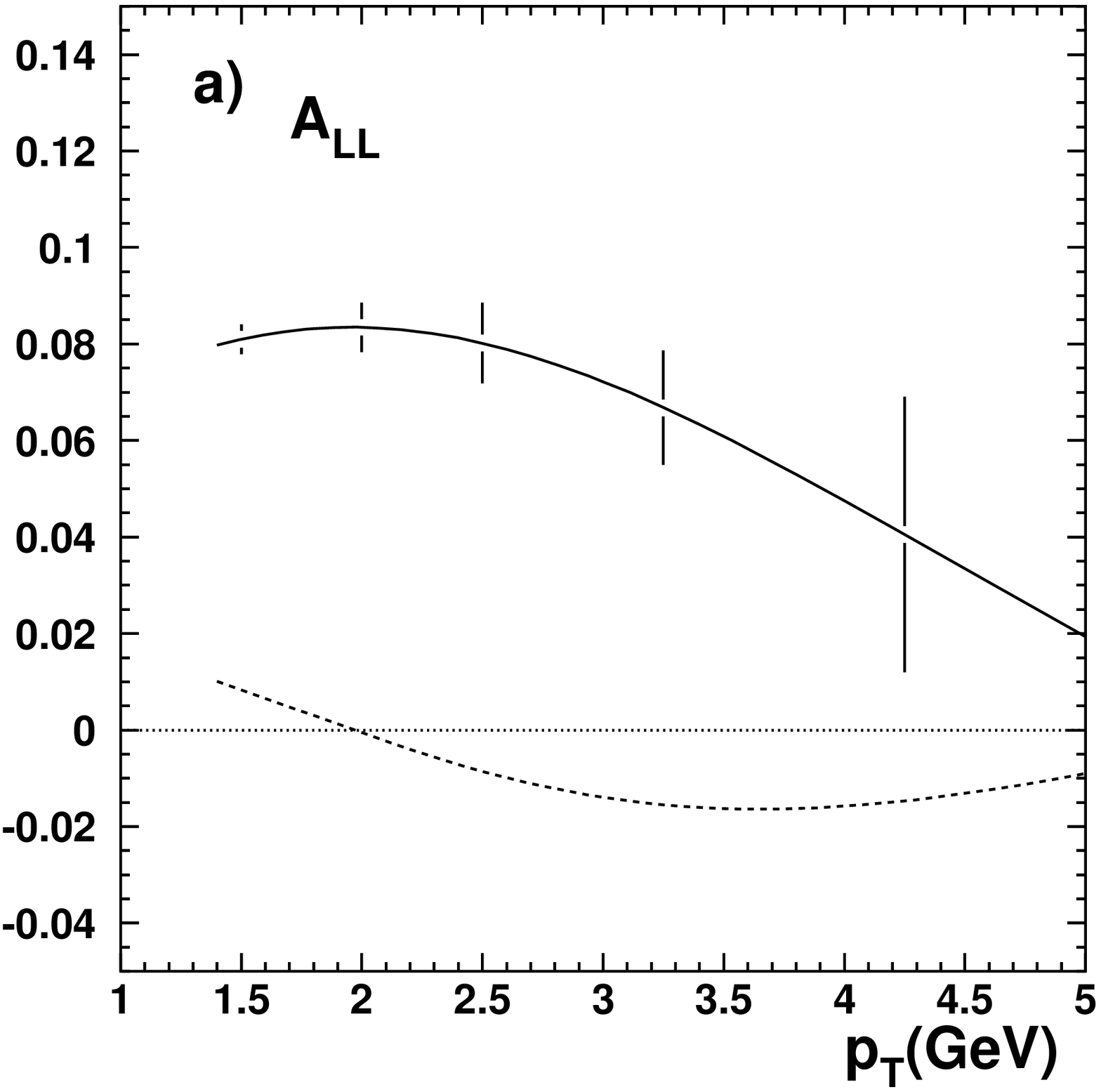, width=8.2cm}
\end{minipage}
\begin{minipage}[c]{8.2cm}
\centering
\epsfig{file=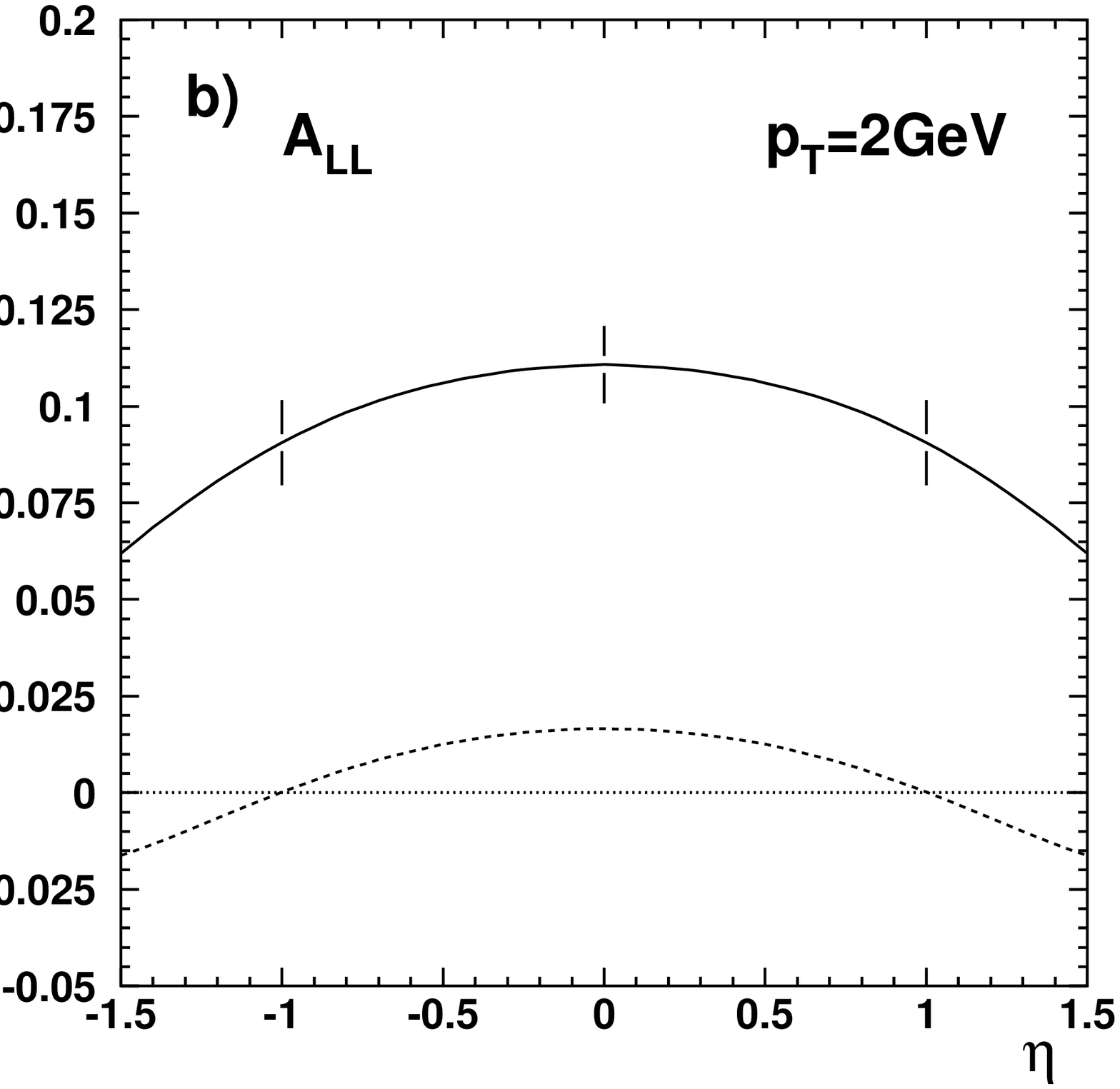,width=8.2cm}
\end{minipage}
\caption{\it Inclusive $J/\psi$ production:
Double spin asymmetry vs. a) $p_T$ and b) $\eta$ for the
LO  set~A (full line) and the set~C (dotted line) of
Ref. \protect \cite{GehrStir_LO}, shown in conjunction with the
HERA--$\vec{N}$ statistical sensitivity. }
         \label{figavto}
\end{figure}
 In fig.~\ref{figavto}a we  present the expected asymmetry versus
 $p_T$ for two different sets
 of polarized gluon distributions taken from \cite{GehrStir_LO}; the 
solid curve  corresponds to  set~A and the dashed one to  set~C.
For  set~A the asymmetry appears sufficiently large
 to be observed and its measurement would allow to extract
information  about the polarized gluon distribution function
in the nucleon.
As can be seen from fig.~\ref{figavto}b, a very good discrimination
between set~A and set~C 
is possible over the whole HERA-$\vec N$ pseudorapidity 
interval.
 For the mass of the charm quark we chose $m_c=1.5$ GeV/c$^2$, as it was  
used for extraction of the color-octet matrix elements from experimental data
\cite{CL,hadro}. We found that 
unlike the $J/\psi$ production cross section the asymmetry
does  practically  not depend on the charm quark mass;
if we vary the charm quark mass from 1.35 to 1.7 GeV/c$^2$ the
magnitude of the asymmetry changes by about $3~\div~5\%$ over the  whole
considered $p_T$ region. \\

{\bf In $J/\psi$ plus jet production} the study of the double spin
asymmetry would allow to access directly the polarized gluon
distribution function, similar to  the case of {\it photon plus jet}
production. For the absolute statistical error of $\Delta G(x_g)/G(x_g)$ 
an expression similar to eq. \ref{err} is obtained:
\begin{eqnarray}
\label{err1}
\delta [{{\Delta G(x_g)}\over {G(x_g)}} ]=
{{\delta A_{LL}}\over {[\Delta G/G] \cdot \hat a_{LL}}}
\end{eqnarray}
\vspace{-1.5em}
with  \\
\vspace{-1.5em}
\begin{eqnarray}
\label{err2}
\hat a_{LL} =
\frac{\Delta\hat\sigma(gg\to J/\psi~g)}
{\hat\sigma(gg\to J/\psi~g)+[q(x)/G(x)]\cdot\hat\sigma(gq\to J/\psi~q)}.
\end{eqnarray}
Here the quark-gluon subprocess was neglected, since its contribution to
$\hat a_{LL}$ amounts only to about 10\% compared to the
gluon-gluon fusion subprocess.  \\
Unfortunately, taking into account the acceptance limitations as it was done
for {\it photon plus jet} production, the $J/\psi$ production
cross section for HERA-$\vec N$  decreases significantly.
Following the same principle of analysis as in Ref. \cite{desy96-04},
it turns out that the measurement of  $\Delta G(x)/G(x)$
in {\it $J/\psi$ plus jet} production is
feasible only for $x_g~=~0.1~\div~0.2$, i.e. for $J/\psi$ transverse
momenta of about 2.5 GeV/c. This prediction is shown as an
additional entry in fig.~\ref{photonjet}. Although being a single point 
only, we underline that this is a very important measurement, because
the lowest lying point from {\it photon plus jet} production is obtained
for rather small values of $p_T$ where perturbative QCD is not
expected to give reliable predictions. We note that the nature of the
gluon-gluon subprocess has rather similar consequences for
{\it jet plus jet} production at RHIC; the prediction
\cite{yok3} consists of only one single point at a similar value of
$x_g$, as well (cf. fig.~\ref{photonjet}).

%
\section{Elastic Scattering}
%

\begin{figure}[ht]
\vspace*{-10mm}
\centering
\epsfig{file=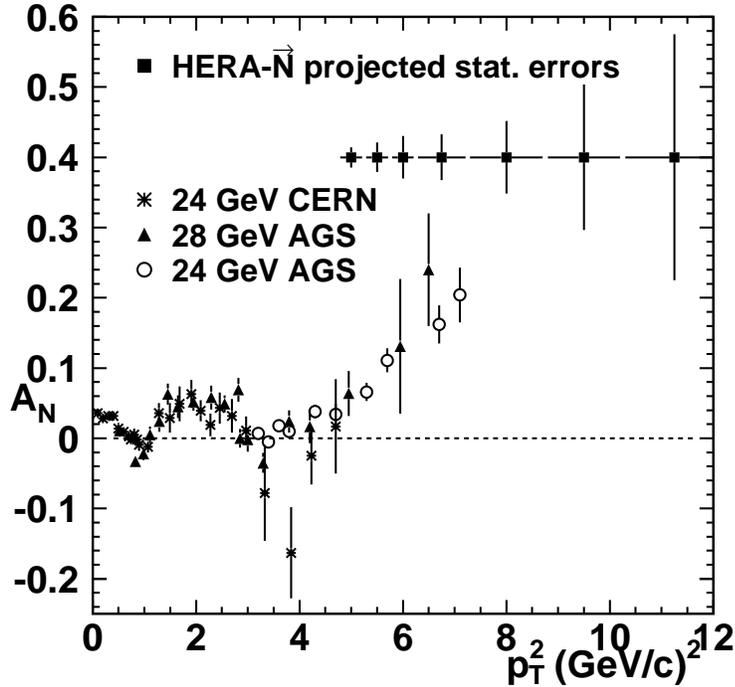,width=11cm}
\caption{\it Compilation of experimental data on the asymmetry in
  elastic proton-proton scattering as a function of $p^2_T$. In
  addition the projected statistical errors attainable with
  HERA-$\vec{N}$ are shown.       } 
    \label{ppelastic}
\vspace*{+10mm}
\end{figure}

Large unexpected spin effects in singly polarized 
proton-proton elastic scattering
$p~+~p^{\uparrow}\rightarrow~p~+~p$ have been discovered many years ago
and remain unexplained up to now. The single spin asymmetry
$A_N$ was found significantly different from zero in the region
$1~\lap~p_T^2~\lap~7$~(GeV/c)$^2$ as it is shown in fig.~\ref{ppelastic} 
in conjunction with the 
HERA-$\vec N$ statistical errors.  
At HERA-$\vec N$ energies
the detection of the recoil proton for 
$p^2_T$ values in the range $5 \div 12$ (GeV/c)$^2$ requires a very large 
angular acceptance (up to 40 degrees) \cite{desy96-04}. 
The forward protons 
for the same interval in $p^2_T$ have laboratory angles of the order
of a few milliradians and require a dedicated detector very 
close to the beam pipe.
Note that although the accessible $p^2_T$ range would be similar
to the range explored at low energies
the c.m. scattering angle detected amounts to a few degrees at
HERA--$\vec{N}$ energies, only.  \\
The transverse single-spin asymmetry $A_N$ in elastic $pp$ scattering 
at HERA-$\vec{N}$ and RHIC energies has been calculated in a
dynamical model that leads to spin-dependent pomeron couplings
\cite{an}. The predicted asymmetry is about 0.1 for 
$p_{T}^2~=~4~\div~5$~(GeV/c)$^2$
with an expected statistical error of $0.01~\div~0.02$ for HERA-$\vec{N}$, i.e.
a significant measurement of the asymmetry $A_N$ can be performed
to test the spin dependence of elastic $pp$ scattering at high energies.

%

\section{Conclusions}
\label{sect5}
%

The physics potential of polarized nucleon-nucleon collisions originating 
from an internal target in the 820 GeV HERA proton beam has been investigated.
Single spin asymmetries, accessible already with the existing unpolarized 
beam, are found to be an almost unique and powerful tool to study the nature
and physical origin of twist-3 effects; even more so when taken 
in conjunction with results of other experiments at HERA. 
When measuring the polarized gluon distribution through double spin 
asymmetries in {\it photon (plus jet)} and {\it $J/\psi$ (plus jet)} 
production -- requiring a polarized HERA proton beam -- the projected 
statistical accuracies are found to be comparable to those predicted for the 
spin physics program at RHIC. In addition,
significant results can be obtained on
the long-standing unexplained spin asymmetries in elastic scattering.
\vspace{+1em}
%

\section*{Acknowledgements}
%

The results presented above were obtained during a several weeks
workshop in Zeuthen which was made possible thanks to the generous
support of DESY-IfH Zeuthen. 
We are indebted to S. Brodsky and W.~Vogelsang for useful discussions and
helpful comments.

\newpage

%

\end{document}